\documentstyle[osa,manuscript]{revtex}

\begin{document}

\begin{center}{\Large \bf Consistency of the Born Approximation for the spin-$\frac{1}{2}$
Aharonov-Bohm Scattering} \\
\vskip.25in
{M\"{u}ge Boz$^{1}$ and  Nam\i{}k K. Pak$^{2}$}

{ $^1$\it Department of Physics Engineering, Hacettepe University, 06532,
Ankara, Turkey }

{ $^2$\it Department of Physics, Middle East Technical University, 06531,
Ankara, Turkey}
 
\end{center}

\begin{abstract}
The relativistic scattering of a spin-$\frac{1}{2}$ particle from an
infinitely long solenoid is considered in the framework of covariant
perturbation theory. The first order term agrees with the corresponding term
in the series expansion of the exact amplitude, and second order term 
vanishes, thus proving that Born approximation is consistent. 

\end{abstract}

\section{INTRODUCTION}
It is by now well-known that when particles of conventional statistics are
coupled to pure Chern-Simons (CS) gauge field, this field creates an
Aharonov-Bohm (AB) like interaction which converts the particles to
charge-flux tube composites \cite{1}. Somewhat later, it was
shown in the context of a Galilean field theory of scalar
fields minimally coupled to a pure CS field that, one can approach 
the problem
of calculating an arbitrary scattering process by restricting
consideration to an N-body sector, allowing  one to derive a 
Schroedinger equation for N-body problem with each pair interacting
as zero radius flux tubes. This has led to the claim that
the field theory, sector by sector,
is formally equivalent to a conventional Schroedinger equation
\cite{2}. Specifically in two particle sector of this equivalent 
field theory, one gets a 
Schroedinger equation similar to the AB equation \cite{3}.

These developments brought back the long-standing issue of failure
of the quantum mechanical perturbation theory for the AB scattering amplitude 
\cite{4}. The
failure of the Born approximation, for instance, is known to be due to the fact 
that the lowest partial wave amplitude
satisfies an integral equation whose interaction term is quadratic in 
terms of the flux parameter. As the exact lowest partial
wave contribution to the scattering amplitude is known to be linear
in this parameter, then,  it is absent in the first order Born
approximation.

There have been several attempts to solve this problem for 
the spinless case  in the context of direct AB
scattering  \cite{5}, anyon physics  
\cite{6}, and  scalar Galilean CS  gauge field theory \cite{7,8,9}.
For instance, in Ref.\onlinecite{7},  
it was shown, through a perturbative 
calculation of the two-particle scattering amplitude, 
up to one-loop order,
that  this amplitude is non-renormalizable, unless a contact 
interaction is introduced, which however for a given strength of the
interaction (critical value corresponding to the self-dual limit)
reduces  to the same order term of the
series expansion of the exact quantum mechanical amplitude.
The same procedure is generalized to the non-Abelian case
with similar conclusions in the second work of  Ref. \onlinecite{7}.
One should note that before the introduction of the contact
interaction, the failure of the naive perturbation expansion of the 
Galilean CS field theory (namely the first Born term for s-wave is
wrong, while the second Born term is infinite) is very similar to that
of the Born series in direct AB scattering. As the exact AB amplitude posseses
scale invariance, it is  natural that the agreement is
obtained  only  after the introduction of contact interaction 
whose coupling strenght
has the critical value for which scale invariance is restored.
This scale invariance at the critical coupling is explicitely checked up to 
three loops  in Ref.\onlinecite{8},  and up to all orders in  
Ref.\onlinecite{9}.

The Born approximation problem for the spinless case was adressed 
from a more general point of view in Ref. \onlinecite{10} and 
Ref.\onlinecite{11} questioning whether the exact (non-perturbative) quantum
mechanical AB amplitude can be reproduced order-by-order
perturbatively in the framework of scalar Galilean CS field
theory. Ref. \onlinecite{10} concludes that the
full agreement is obtained if the  renormalized strength of the
contact interaction is chosen to be related to the
self-adjoint extension parameter, for fixed renormalization scale.
However, the conclusion of more recent work  \cite{11} is 
not in full
agreement with that of  the  Ref.\onlinecite{10}. They 
show that the full 
agreement can be obtained only in some special regimes.
Thus, we see that the general problem in the context of Galilean scalar
field theory is not satisfactorily settled yet.
 
In Ref.\onlinecite{12}, it was shown that if one starts from the 
relativistic scalar
gauge field theory of the CS interaction, one finds a renormalizable
one-loop scattering amplitude which remains so in the non-relativistic
limit as well, thus reproducing the correct series expansion of the 
exact quantum mechanical expression without the need to introduce
a contact interaction term. It is not clear yet whether the issue 
raised in Ref.\onlinecite{10} and Ref.\onlinecite{11} would be relevant 
for the relativistic
field theories. Obviously there are some fundamental differences
between the non-relativistic and the relativistic cases. For instance,
in the non-relativistic case, the necessity of a cut-off is not a
relic of some unknown ultraviolet physics, but rather an artefact of
the perturbative methods used. This is in contrast with the conventional
wisdom on renormalization, whose natural habitat is the relativistic
field theories.

AB scattering of spin-$\frac{1}{2}$ particles from an infinitely
long solenoid was considered in the context of Dirac equation formalism
in Ref. \onlinecite{13}, and using covariant perturbation theory in 
Ref. \onlinecite{14}. In
these works, it was shown that
that Born approximation works, that is, it agrees with the
corresponding term in the series expansion of the exact amplitude.
The agreement obtained in the framework of Dirac equation is not 
surprising at all. Because the Dirac Hamiltonian is linear in momenta,
and the corresponding integral equation determining the lowest partial 
wave amplitude involves a term  linear in flux parameter.

The spin-$\frac{1}{2}$ AB problem was also considered in the
framework of equivalent Galilean CS gauge field theory in Ref. \onlinecite{15}
and Ref. \onlinecite{16} from  different perspectives. In these works
the consistency of the perturbative treatment was checked
up to one-loop order. As the exact amplitude is proportional
to $\sin\pi\alpha$ (with $\alpha=-\frac{e \phi}{2 \pi}$,
and $\phi$ is the magnetic flux carried by the solenoid),
the series expansion of this term contains terms of order
O($\alpha$), O($\alpha^{3}$),{\ldots};  that is O($\alpha^{2}$)
is missing. Thus a complete check of the consistency 
of perturbative
approach, not only should get agreement on O($\alpha$)
terms, but also should
show the vanishing of  O($\alpha^{2}$)  terms  
(1-loop terms in the language of the 
field theory).

In Ref. \onlinecite{15}, it was shown that the two-particle sector of the Galilean 
field theory again leads to an AB-like equation. Then, the two particle
scattering amplitude is computed up to 1-loop order. The tree-contribution
(O($\alpha$)) agrees with the corresponding term in the series 
expansion of the exact amplitude; the 1-loop contribution
(O($\alpha^{2}$)) is finite and vanishes. This completes the check 
of consistency of the Born approximation to lowest order, in the
(sector by sector) equivalent field theory framework.

Encouraged by the results of Ref. \onlinecite{15} and 
Ref. \onlinecite{16}  in the Galilean field 
theory framework, it is aimed in this work to carry out the 
second order analysis in direct version of the problem,
namely the relativistic scattering of spin-$\frac{1}{2}$
particles from an infinitely long solenoid, and check the
consistency of the Born approximation fully, by demonstrating that
O($\alpha^{2}$) contribution to the scattering amplitude in the
framework of covariant perturbation theory vanishes. We will show 
that this is indeed what happens.

We would like to note that the subtleties pointed out in
Ref.\onlinecite{10} and Ref.\onlinecite{11} for the spinless case 
are natually overcome
in the relativistic case considered in this work. This does not
create any difficulty in establishing parallelism with the
results obtained in Ref.\onlinecite{15} in the context of 
Galilean CS field theory. Because it was already shown in 
Ref.\onlinecite{15} that, in contrast with the crucial role
played by the contact interaction in the scalar case, the
contribution of the Pauli term formally corresponding to the 
contact interaction  (produced
in the non-relativistic limit of the fermionic CS gauge field
theory with given coupling strength) to 1-loop diagrams are
finite and null, thanks to the statistics.

This paper is organized as follows:
In Sect $2$, we briefly review the results of Ref. \onlinecite{14} for the general
discussion of the Helicity conservation. In Sect $3$, we review
the covariant perturbation theory approach to lowest order
for the problem under consideration. In Sect $4$, the O($\alpha^{2}$)
contribution to the scattering amplitude is computed; and it
is shown that this contribution vanishes. Sect $5$ is devoted to the discussion of
the results.

\section{HELICITY CONSERVATION AND THE EXACT SCATTERING AMPLITUDE}

The basic starting point of Ref. \onlinecite{14} is the well-known observation
that the helicity of a spin-$\frac{1}{2}$ particle is
unchanged by a time-independent magnetic field \cite{17}.

Defining the Helicity eigenstates in the initial and final
states as $|\pm\rangle_{i,f}$ , the first observation is that
$|\pm\rangle_{i}\rightarrow|\pm\rangle_{f}$ transitions
proceed with unit probability in the Helicity space. Denoting the
scattering matrix by S this reads as
\begin{eqnarray}
|_{f}\langle\pm|S|\pm\rangle_{i}|^{2}=1 \nonumber\\
|_{f}\langle\pm|S|\mp\rangle_{i}|^{2}=0
\label{eq-1}
\end{eqnarray}
Thus the differential cross section for $|\pm\rangle_{i}\rightarrow|\pm\rangle_{f}$ per unit length is 
determined by the phase space only, and thus equal to the
unpolarized cross section: 

We next consider the scattering from an initial state polarized
along the direction of an arbitrary unit vector $\hat{n}$
to a final state moving along $\theta$, in which the beam is
polarized again in the same $\hat{n}$
direction. Denoting the spherical angles of $\hat{n}$ with respect
to the initial beam axis (chosen as x-axis) by
($\theta', \varphi'$  )
these states are given as
\begin{eqnarray}
|i(\vec{p}_{i},\hat{n})\rangle&=&\cos\frac{\theta'}{2}e^\frac{-i\varphi'}{2}|+\rangle_{i}+\sin\frac{\theta'}{2}e^\frac{i\varphi'}{2}|-\rangle_{i}
\nonumber\\
|f(\vec{p}_{f},\hat{n})\rangle&=& 
(\cos\frac{\theta}{2}\cos\frac{\theta'}{2}e^\frac{-i\varphi'}{2}+
\sin\frac{\theta}{2}\sin\frac{\theta'}{2}e^\frac{i\varphi'}{2}|+\rangle_{f}\nonumber\\
&&
+(\cos\frac{\theta}{2}\sin\frac{\theta'}{2}e^\frac{i\varphi'}{2}-
\sin\frac{\theta}{2}\cos\frac{\theta'}{2}e^\frac{-i\varphi'}{2}|-\rangle_{f}
\label{eq-2}
\end{eqnarray}
Using equations ~(\ref{eq-2}), one readily gets
\begin{eqnarray}
\langle f|S|i\rangle=\cos\frac{\theta}{2}-i\sin\frac{\theta}{2}\sin\theta'
\sin\varphi'
\label{eq-3}
\end{eqnarray}

Thus the polarized cross section per unit length of the solenoid
is obtained as
\begin{eqnarray}
\frac{d\sigma}{d\theta}=[1-(\hat{n}\times\hat{z})^{2}\sin^{2}\frac{\theta}{2}]
(\frac{d\sigma}{d\theta})_{unpol}
\label{eq-4}
\end{eqnarray}
Here $\hat{z}$ is the unit vector in the direction of the 
solenoid. Thus the cross section differs from the unpolarized case
(or the spinless case) when the spin of the particle has
components in the scattering plane(chosen here as x-y plane).

\section{COVARIANT PERTURBATION THEORY; FIRST ORDER (BORN APPROXIMATION)}
The purpose of this section is to show that Born approximation 
reproduces the correct result, that is it agrees with the 
corresponding terms in the series expansion of the 
exact amplitude. 

The S-matrix element for a spin-$\frac{1}{2}$
particle scattering from an external electromagnetic field
to lowest order is given by:
\begin{eqnarray}
S^{(1)}_{fi}=\int d^{4}z\bar{\psi}_{f}(z)(ie\gamma_{\mu}A^{\mu}(z))\psi_{i}(z)
\label{eq-5}
\end{eqnarray}
where in the Bjorken-Drell convention
\begin{eqnarray}
\psi_{i}(z)&=&\sqrt{\frac{m}{E_{i}V}}u(p_{i},s_{i})e^{-ip_{i\mu}z^{\mu}}\nonumber\\
\psi_{f}(z)&=&\sqrt{\frac{m}{E_{f}V}}u(p_{f},s_{f})e^{-ip_{f\mu}z^{\mu}}
\label{eq-6}
\end{eqnarray}
The vector potential of the solenoid, taken along the 3rd axis,
in the Coulomb gauge $\vec{\nabla}\cdot \vec{A}=0$
is given as
\begin{eqnarray}
A_{1}(z)&=&-\frac{\phi}{2\pi}\frac{z_{2}}{z^{2}_{1}+z^{2}_{2}} \nonumber\\
A_{2}(z)&=&-\frac{\phi}{2\pi}\frac{z_{1}}{z^{2}_{1}+z^{2}_{2}} \nonumber\\
A_{3}(z)&=&A_{0}(z)=0
\label{eq-7}
\end{eqnarray}
where $\phi$ is the magnetic flux carried by the solenoid. Denoting
$\vec{q}=\vec{p_{f}}-\vec{p_{i}}$, and carrying out the z-
integrals, we find

\begin{eqnarray}
S^{(1)}_{fi}=\frac{4\pi^{2}}{V}(me\phi)\delta(E_{f}-E_{i})\delta(p_{f3}-p_{i3})
\frac{\bar{u}(f)(\gamma^{2}q_{1}-\gamma^{1}q_{2})u(i)}{\sqrt{E_{f}E_{i}}(q^{2}_{1}+q^{2}_{2})}
\label{eq-8}
\end{eqnarray}
As the initial beam is in the 1st direction ($p_{i3}=0$),
denoting $t=\bar{u}(f)(\gamma^{2}q_{1}-\gamma^{1}q_{2})u(i)$,
the differential cross section per unit solenoid length,
to this order, is given as
\begin{eqnarray}
(\frac{d\sigma}{d\theta})^{Born}=\frac{m^{2}e^{2}\phi^{2}}{2\pi|
\vec{p_{i}}|(q_{1}^{2}+q_{2}^{2})^{2}}|t|^{2}
\label{eq-9}
\end{eqnarray}
with $|\vec{p}_{i}|=|\vec{p}_{f}|=k$ and $E_{i}=E_{f}$,
as imposed by the $\delta$-functions. We can proceed in
two ways:
a)We can sum over final polarizations, and average over
the initial ones  to get the unpolarized cross section 
by direct use of  Dirac matrix algebra
\begin{eqnarray}
(\frac{d\sigma}{d\theta})^{Born}=\frac{e^{2}\phi^{2}}{8\pi k \sin^{2}\frac{\theta}{2}}
\label{eq-10}
\end{eqnarray}
where $ \vec{p_{i}}=k\hat{x}$.
b)We can compute the polarized amplitude, using the 
explicit expressions of the Dirac spinors for the polarized
initial and final electrons.
\begin{eqnarray}
u(i)&=&\cos\frac{\theta'}{2}e^\frac{-i\varphi'}{2}u_{+}(i)+\sin\frac{\theta'}{2}e^\frac{i\varphi'}{2}u_{-}(i)
\nonumber\\
u(f)&=&(\cos\frac{\theta}{2}\cos\frac{\theta'}{2}e^\frac{-i\varphi'}{2}+
\sin\frac{\theta}{2}\sin\frac{\theta'}{2}e^\frac{i\varphi'}{2})u_{+}(f)\nonumber\\
&&
+(\cos\frac{\theta}{2}\sin\frac{\theta'}{2}e^\frac{i\varphi'}{2}-
\sin\frac{\theta}{2}\cos\frac{\theta'}{2}e^\frac{-i\varphi'}{2})u_{-}(f)
\label{eq-11}
\end{eqnarray}
where
\begin{eqnarray}
u _{+}(i)&=&N_{i} \left( \begin{array}{c}
1\\
0\\
\mu _{i}\\
0
\end{array}\right),\,\,\,\,\,  u _{-}  (i)=N_{i}\left( \begin{array}{c}
0\\
1\\
0\\
-\mu _{i}
\end{array}\right)   
\nonumber\\
N_{i}&=&\sqrt{\frac{E_{i}+m}{2m}},\,\,\,\,\,  \mu_{i}=\frac{|\vec{p}_{i}|}{E_{i}+m}
\nonumber\\
u _{+}(f)&=&N _{f}\left( \begin{array}{c}
cos\frac{\theta }{2}\\
sin\frac{\theta }{2}\\
\mu _{f}cos\frac{\theta }{2}\\
\mu _{f}sin\frac{\theta }{2}
\end{array}\right),\,\,\,\,\,  u _{-} (f)=N_{f}\left( \begin{array}{c}
-sin\frac{\theta }{2}\\
cos\frac{\theta }{2}\\
\mu _{f}sin\frac{\theta }{2}\\
-\mu _{f}cos\frac{\theta }{2}
\end{array}\right)  
\nonumber\\ 
N_{f}&=&\sqrt{\frac{E_{f}+m}{2m}},\,\,\,\,\,  \mu_{f}=\frac{|\vec{{p}_{f}}|}{E_{f}+m}
\label{eq-12}
\end{eqnarray}
Using ~(\ref{eq-11}), ~(\ref{eq-12}), we can compute t, and find
\begin{eqnarray}
t=-\frac{2k^{2}}{m} \sin\frac{\theta}{2}[\cos\frac{\theta}{2}-
i\sin\frac{\theta}{2}\sin\theta'\sin\varphi']
\label{eq-13}
\end{eqnarray}
Substituting ~(\ref{eq-13}) in ~(\ref{eq-9}), we get
\begin{eqnarray}
(\frac{d\sigma}{d\theta})_{pol}^{Born}=
(\frac{d\sigma}{d\theta})_{unpol}^{Born}
[1-(\hat{n} \times \hat{z})^{2} \sin^{2}\frac{\theta}{2}]
\label{eq-14}
\end{eqnarray}
Thus, Born approximation indeed works in the
polarized case. The scattering amplitude (and thus the cross section)
is effected by the same expression in
the Born approximation as in  case of the exact amplitude.
However this does not constitute a complete check of the
consistency of the Born approximation in the relativistic
spin-$\frac{1}{2}$ AB effect yet. As the exact amplitude is
proportional to $\sin\pi\alpha$, a full consistency
would require that the $O(\alpha^{2})$
contribution to the scattering amplitude should vanish; and
this is what we will check next.

\section{COVARIANT PERTURBATION THEORY- SECOND ORDER}
The S-matrix  in the second order is given as
\begin{eqnarray}
S^{(2)}_{fi}  = \int \int d^{4}xd^{4}y\bar{\psi _{f}}(x)
(-ie \gamma^{\mu }A_{\mu }(x) )iS_{F}(x-y) 
(-ie \gamma ^{\nu }A_{\nu } (y))\psi _{i}(y)
\label{eq-15}
\end{eqnarray}
where 
$S_{F}(x-y)=\int \frac{d^{4}p}{(2\pi)^{4}}e^{-ip(x-y)}
\frac{\gamma^{\mu}p_{\mu}+m}{p^{2}-m^{2}+i\varepsilon }.$
Carrying out the spatial integrals we get
\begin{eqnarray}
S_{fi}^{(2)}&=&\frac{i}{V}(e^{2}\phi ^{2})\frac{m}{\sqrt{E_{i}E_{f}}}\delta (E_{f}-E_{i})\delta (p_{f3}-p_{i3})I
\nonumber\\
I&=&\int d^{2}p_{\perp}\frac{N}
{(\vec {p_{f_{ \perp}}}^{2}-\vec {p_{\perp}}^{2})
(\overrightarrow{p_{f}-p})_{\perp}^{2}(\overrightarrow{p_{i}-p})_{\perp}^{2}}
\label{eq-16}
\end{eqnarray}
where N is obtained as
\begin{eqnarray}
N&=&(p_{i}-p)_{2}(p_{f}-p)_{1}\bar{u}_{f}\gamma^{1}P\gamma^{3}u_{i}+
(p_{i}-p)_{1}(p_{f}-p)_{2}\bar{u}_{f}\gamma^{3}P\gamma^{1}u_{i} \nonumber\\
&&
-(p_{i}-p)_{1}(p_{f}-p)_{1}\bar{u}_{f}\gamma^{1}P\gamma^{1}u_{i}-
(p_{i}-p)_{2}(p_{f}-p)_{2}\bar{u}_{f}\gamma^{3}P\gamma^{3}u_{i}
\end{eqnarray}
with 
\begin{eqnarray}
P=\gamma^{0}E_{f}-\gamma^{3}p_{1}-\gamma^{1}p_{2}+m
\label{eq-17}
\end{eqnarray}
Denoting the polar angle in the $p_{\perp}$ plane by
$\varphi$, and making use of the energy conservation mandated
by $\delta(E_{f}-E_{i}),\,\,\,\,\, \vec{p_{i}}^{2}=\vec{p_{f}}^{2}\equiv k^{2}$
with $\vec{p}_{i}=k\hat{x}$, then N can be written as
\begin{eqnarray}
N&=&\alpha+\beta\cos\varphi+\gamma\sin\varphi \nonumber\\
\alpha&=&k^{3}\{\frac{E_{i}}{k}\{ A\sin\theta-B\cos\theta\}
-u^{2}\{\frac{E_{i}}{k}B+D\sin\theta+C(1+\cos\theta)\} \nonumber\\
&& 
+ \frac{m}{k}\{ A'\sin\theta+B'\cos\theta+B'u^{2}\} \} \nonumber\\
\beta&=&k^{3}\{Cu^{3}+u\{ D\sin\theta+C\cos\theta+
\frac{E_{i}}{k}\{(1+\cos\theta)
B-A\sin\theta)\}\} \nonumber\\
&& 
-\frac{mu}{k}\{ A'\sin\theta+B'(1+\cos\theta)\}\} \nonumber\\
\gamma&=&k^{3}\{Du^{3}+u\{ C\sin\theta-D\cos\theta
-\frac{E_{i}}{k}\{ A(1-\cos\theta)-B\sin\theta\}\} \nonumber\\
&&
-\frac{mu}{k}\{ A'(1-\cos\theta)+B'\sin\theta\}\}
\label{eq-18}
\end{eqnarray}
with $u\equiv\frac{p}{k}$ and 
\begin{eqnarray}
A&=&i\bar{u}_{f}\gamma^{0}{\mit\Sigma}_{2} u_{i},\,\,\,\, 
A'=-i\bar{u}_{f}{\mit\Sigma}_{2}u_{i},\,\,\,\,\,\,
\rm with \,\,\,\,\,\,\,
{\mit\Sigma}_{2}=\left( \begin{array}{cc}
\sigma_{2} & 0\\
0 & \sigma_{2}
\end{array}\right) \nonumber \\
B&=&\bar{u}_{f}\gamma^{0}u_{i},\,\,\,\,\,B'=\bar{u}_{f}u_{i}\nonumber\\
C&=&\bar{u}_{f}\gamma^{3}u_{i},\,\,\,\,\,\,D=\bar{u}_{f}\gamma^{1}u_{i}
\label{eq-19}
\end{eqnarray}

The $\varphi$ integration can be carried out using the 
complex integration techniques. That is we define 
$z=e^{i\varphi}$, and the $\varphi $ integration is
converted into a contour integration over the unit circle
$|z|=1$. Thus
\begin{eqnarray}
I=\frac{e^{i\theta}}{2ik} \int_{0}^{\infty} \frac{du}{u(1-u^{2})}
\oint_{|z|=1}dz\frac{{\cal F}(z, \bar{z})}
{(z^{2}+1-2az)(z^{2}+e^{2i\theta}-2aze^{i\theta})}
\label{eq-20}
\end{eqnarray}
with
$a=\frac{u^{2}+1}{2u}$ and
\begin{eqnarray}
{\cal F}(z, \bar{z})=c_{0}+c_{1}z+c_{2}z^{2}
\end{eqnarray}
where
\begin{eqnarray}
c_{0}&=&\{ C+iD\} u^{3}+ u\{(C-iD)e^{i\theta}+
\frac{E_{i}}{k}(B-iA +(B+iA)e^{i\theta})\nonumber\\
&&
-\frac{m}{k}( B'+iA'+(B'-iA')e^{i\theta})\} \nonumber\\
c_{1}&=&\frac{2E_{i}}{k}\{ A\sin\theta-B\cos\theta\}
-2u^{2}\{\frac{E_{i}}{k} B -\frac{m}{k}B'+ D \sin\theta+ 
C(1+\cos\theta)\} \nonumber\\
&&
+2\frac{m}{k}\{ A'\sin\theta+B'\cos\theta\}\nonumber\\
c_{2}&=&\{ C-iD\}  u^{3}+u\{(C+iD)e^{-i\theta}+
\frac{E_{i}}{k}(B+iA+(B-iA)e^{-i\theta})\nonumber\\
&&
-\frac{m}{k}( B'-iA'+(B'+iA')e^{-i\theta})\}
\label{eq-21}
\end{eqnarray}
The z-integral now can be carried out, using the Cauchy 
theorem, and we get
\begin{eqnarray}
J&=&-2\pi i e^{-i\theta}\frac{2u^{2}}{k}\nonumber\\
&&
\times \frac{\{(E_{i}B-mB')u^{2}+E_{i}(A\sin\theta-B\cos\theta)+m
(A'\sin\theta+B'\cos\theta)\}}
{(u^{2}-e^{i\theta})(u^{2}-e^{-i\theta})} \varepsilon(u-1)
\label{eq-22}
\end{eqnarray}
Substituting ~(\ref{eq-22}) in ~(\ref{eq-20}) , we get
\begin{eqnarray}
I&=&\frac{2\pi}{k^{2}}\int_{0}^{\infty}udu\varepsilon(u-1)\nonumber\\
&&
\times\frac{\{(E_{i}B-mB')u^{2}+E_{i}(A\sin\theta-B\cos\theta)+m
(A'\sin\theta+B'\cos\theta)\}} 
{(u^{2}-1)(u^{2}-e^{i\theta})(u^{2}-e^{-i\theta})}
\label{eq-23}
\end{eqnarray}
Changing variables, $u^{2}=v$, ~(\ref{eq-23}) could be rewritten as
\begin{eqnarray}
I&=&
\frac{\pi}{k^{2}}\int_{0}^{\infty} dv \varepsilon(v-1)\nonumber\\
&&
\times \{\frac{E_{i}B-mB'}{(v-e^{i\theta})(v-e^{-i\theta})}+ 
\frac{(E_{i}A+mA')\sin\theta+(E_{i}B-mB')(1-\cos\theta)}
{(v-1)(v-e^{i\theta})(v-e^{-i\theta})}\}
\label{eq-24}
\end{eqnarray}
The first integral in ~(\ref{eq-24}), can easily be shown to vanish, with
the help of  a variable change $v=\frac{1}{w}$ in the
$(1,\infty)$ interval. Thus, we finally end up with
\begin{eqnarray}
I= \frac{\pi T}{k^{2}}\int_{0}^{\infty} dv \frac{\varepsilon(v-1)}
{(v-1)(v-e^{i\theta})(v-e^{-i\theta})} 
\label{eq-25}
\end{eqnarray}
where
\begin{eqnarray}
T&=&(E_{i}A+mA')\sin\theta+(E_{i}B-mB')(1-\cos\theta)\nonumber\\
&=&\bar u_{f}(E_{i}\gamma^{0}-m)(1-\cos\theta+i
\sin\theta {\mit\Sigma}_{2})u_{i}
\end{eqnarray}
Using the definition in~(\ref{eq-19}), 
the profactor T can be shown to vanish. 

\section{CONCLUSIONS AND DISCUSSION}
In Ref. \onlinecite{14}  it was claimed that the Born approximation for relativistic
spin-$\frac{1}{2}$ AB scattering works, by demonstrating that
this amplitude agrees with the corresponding terms in the series
expansion of the exact amplitude. As the exact amplitude is proportional
to $\sin\pi\alpha$, the demonstration of the full consistency of
the Born approximation however requires a further step, namely the
vanishing of the $O(\alpha^{2})$ contributions. This was already done
in the context of the Galilean invariant field theory whose 2-particle
sector is known to be equivalent to the AB Schroedinger equation.
Encouraged by the success of these works, we have addressed the same
issue directly, namely by considering the $O(\alpha^{2})$ contribution
for the relativistic scattering of spin-$\frac{1}{2}$ particles from an
infinitely long solenoid in the context of covariant perturbation theory,
and shown that it indeed vanishes, thus completing the consistency
check of the Born approximation for the relativistic spin-$\frac{1}{2}$
problem.

\end{document}